\documentclass[a4paper,11pt]{article}
\usepackage{pos}
\usepackage{booktabs}
\usepackage{mathtools}

\newcommand{\rmd}{\mathrm{d}}
\newcommand{\rme}{\mathrm{e}}
\newcommand{\rmi}{\mathrm{i}}
\DeclareMathOperator{\re}{Re}

\DeclareMathOperator{\tr}{tr}

\newcommand{\calZ}{\mathcal{Z}}
\newcommand{\SW}{S_{\mathrm{W}}}
\newcommand{\Qtop}{Q_{\mathrm{top}}}
\newcommand{\tflow}{t_{\mathrm{flow}}}
\newcommand{\Ns}{N_{\mathrm{s}}}
\newcommand{\Nt}{N_{\mathrm{t}}}
\newcommand{\Nconf}{N_{\mathrm{conf}}}
\newcommand{\avg}[1]{\left\langle #1\right\rangle}

\title{Dyon Condensation at $\theta=2\pi$ from Wilson--'t~Hooft Loops in Yang--Mills Theory}
\ShortTitle{Dyon Condensation at $\theta=2\pi$ from Wilson--'t~Hooft Loops in YM Theory}

\author[a,b]{Hiromasa Watanabe}
\author[c]{Issaku Kanamori}
\author*[d]{Okuto Morikawa}
\author[e,f]{Yuki Nagai}
\author[a]{Yuya Tanizaki}
\author[g,h,c]{Akio Tomiya}

\affiliation[a]{Yukawa Institute for Theoretical Physics, Kyoto University,
Kitashirakawa Oiwakecho, Sakyo-ku, Kyoto 606-8502, Japan}
\affiliation[b]{Department of Physics, and Research and Education Center for
Natural Sciences, Keio University,
4-1-1 Hiyoshi, Yokohama, Kanagawa 223-8521, Japan}
\affiliation[c]{RIKEN Center for Computational Science,
Kobe 650-0047, Japan}
\affiliation[d]{Center for Interdisciplinary Theoretical and Mathematical
Sciences (iTHEMS), RIKEN, Wako 351-0198, Japan}
\affiliation[e]{Information Technology Center, The University of Tokyo,
Kashiwa, Chiba 277-0882, Japan}
\affiliation[f]{Department of Advanced Materials Science,
The University of Tokyo, Kashiwa, Chiba 277-8561, Japan}
\affiliation[g]{Department of Information and Mathematical Sciences,
Tokyo Woman's Christian University, Tokyo 167-8585, Japan}
\affiliation[h]{Department of Physics, Kyoto University,
Kyoto 606-8502, Japan}

\emailAdd{hiromasa.watanabe@yukawa.kyoto-u.ac.jp}
\emailAdd{okuto.morikawa@riken.jp}

\abstract{%
Although Yang--Mills theories at $\theta$ and $\theta+2\pi$ are physically
equivalent, their periodicity is realized nontrivially in the presence of a
background field for the $1$-form center symmetry.  The confining vacua at
$\theta=0$ and $2\pi$ are therefore expected to have distinct
symmetry-protected topological labels.  We test this expectation in
$SU(2)$ lattice Yang--Mills theory by measuring 't~Hooft and dyonic
Wilson--'t~Hooft loops at $\theta=2\pi$.  A $1$-form-covariant DBW2 flow
makes the global reweighting factor numerically benign while retaining the
physical non-integer topological charge in a defect ensemble.  The dyonic
loop exhibits a clear perimeter-law plateau on two lattice volumes.  The
direct 't~Hooft-loop signal is noise-limited but has the magnitude and
distance dependence expected for an area law.  Taken together with the Wilson--'t~Hooft classification, these results provide numerical evidence for the dyonic oblique-confinement pattern at $\theta=2\pi$.
This talk is based on our recent paper~\cite{Watanabe:2026ymb}.}

\FullConference{The 43rd International Symposium on Lattice Field Theory
(Lattice 2026)\\
July 26 to August 1, 2026\\
University of Maryland, College Park, USA}

\begin{document}
\maketitle
\flushbottom

\section{Introduction}
\label{sec:introduction}

Confinement in four-dimensional Yang--Mills theory can be characterized by
the area law of a fundamental Wilson loop~\cite{Wilson:1974sk}.  The
$\theta$ parameter gives a particularly useful way to probe the 
screening structure of the confining vacuum.  By the Witten effect, an object carrying
electric and magnetic charges $(e,m)$ has the effective electric charge
\begin{equation}
  e_{\mathrm{eff}}=e+\frac{\theta}{2\pi}m .
  \label{eq:witten_effect}
\end{equation}
Thus a monopole with $(e,m)=(0,1)$ at $\theta=0$ becomes a dyon with
$(e,m)=(1,1)$ after shifting $\theta$ by $2\pi$~\cite{Witten:1979ey}.

This conclusion does not rely on the semiclassical dual-superconductor
picture.  Couple the $\mathbb{Z}_N$ $1$-form center
symmetry of pure $SU(N)$ Yang--Mills theory to a background two-form gauge
field $B$.  The partition function then obeys
\begin{equation}
  \calZ_{\theta+2\pi}[B]
  =
  \exp\left[
    -\frac{2\pi\rmi}{N}
    \int_{T^4}\frac{1}{2}B\mathbin{\cup}B
  \right]
  \calZ_{\theta}[B] .
  \label{eq:theta_anomaly}
\end{equation}
If the center symmetry remains unbroken, the phase in
Eq.~\eqref{eq:theta_anomaly} shifts the discrete symmetry-protected
topological (SPT) label of the gapped vacuum~\cite{Gaiotto:2017yup}.
Consequently, at least one phase transition must separate the vacua at
$\theta=0$ and $2\pi$.

The Wilson--'t~Hooft classification turns this statement into a
gauge-invariant numerical test~\cite{tHooft:1977nqb,tHooft:1979rtg}.
The Wilson loop $W(C)$ probes electric charge, the 't~Hooft loop
$H(\widetilde C,\widetilde\Sigma)$ probes magnetic charge, and their product
$HW$ probes a dyon.  A screened probe has a perimeter law, whereas an
unscreened probe has an area law.  For a confining $SU(2)$ vacuum, the two
possibilities relevant here are summarized in Table~\ref{tab:classification}.
The Wilson area law is already guaranteed by confinement.  We therefore test
the two starred entries at $\theta=2\pi$, following the detailed study in
Ref.~\cite{Watanabe:2026ymb}.

\begin{table}[h]
  \centering
  \small
  \begin{tabular}{@{}lcccc@{}}
    \toprule
    Condensate
      & $\avg{W}$
      & $\avg{H}$
      & $\avg{HW}$
      & SPT label \\
    \midrule
    Monopole & Area & Perimeter & Area & Trivial \\
    Dyon     & Area & Area$^\star$ & Perimeter$^\star$ & Nontrivial \\
    \bottomrule
  \end{tabular}
  \caption{Wilson--'t~Hooft classification of the two confining phases
  relevant to this work.  The starred behaviors distinguish the
  $\theta=2\pi$ vacuum from the monopole-condensed vacuum at $\theta=0$.}
  \label{tab:classification}
\end{table}

Here ``dyon condensation'' is used operationally for the
Wilson--'t~Hooft phase in which the $(e,m)=(1,1)$ line is screened; it does
not require a gauge-dependent identification of a local dyon condensate.

\section{Lattice defect and reweighting strategy}
\label{sec:method}

\subsection{The 't~Hooft defect}

\begin{figure}[t]
  \centering
  \includegraphics[width=0.8\linewidth]{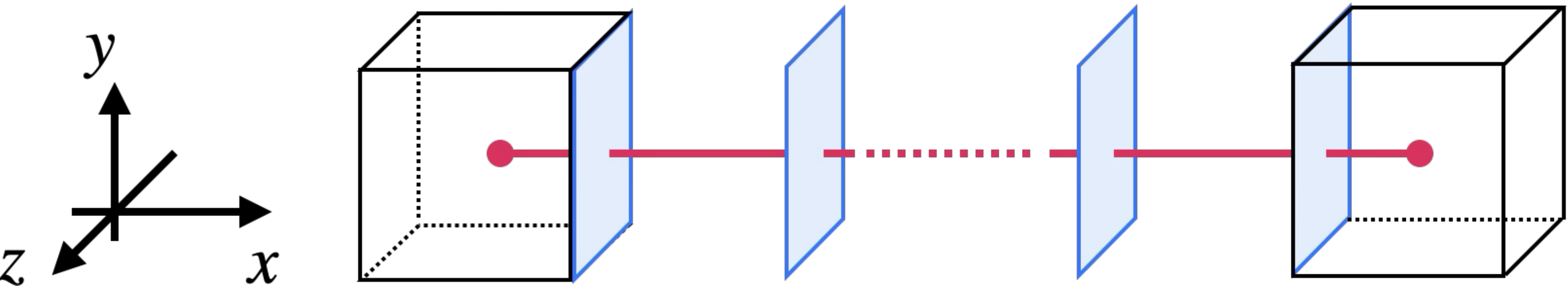}
  \caption{Magnetic probes on a time slice of the dual lattice.  The
  background field $B$ is activated on the plaquettes pierced by the
  Dirac sheet.  The endpoints are separated by $R$. Reproduced from Ref.~\cite{Watanabe:2026ymb}.}
  \label{fig:defect}
\end{figure}

Let $U_\ell\in SU(2)$ be the link variable and
$B_p\in\mathbb{Z}_2$ a background plaquette field.  We use the Wilson action
\begin{equation}
  \SW[U,B]
  =
  -\frac{\beta}{2}
  \sum_p (-1)^{B_p}\re\tr U_p ,
  \label{eq:wilson_action}
\end{equation}
which is invariant under both ordinary $SU(2)$ gauge transformations and
background $\mathbb{Z}_2$ $1$-form gauge transformations
~\cite{Gaiotto:2014kfa}.  An open surface
$\widetilde\Sigma$ on the dual lattice specifies
\begin{equation}
  B=\delta[\widetilde\Sigma],
  \qquad
  \rmd B=\delta[\widetilde C],
  \qquad
  \partial\widetilde\Sigma=\widetilde C .
  \label{eq:defect}
\end{equation}
The non-flatness at $\widetilde C$ inserts a magnetic probe and defines the
't~Hooft loop.  We choose an $R\times\Nt$ surface wrapping the Euclidean-time
circle; its boundary represents a static monopole--antimonopole pair
separated by $R$, as shown in Fig.~\ref{fig:defect}.

\subsection{Factorization at finite \texorpdfstring{$\theta$}{theta}}

For an electric observable $\mathcal{O}$, the defect expectation value can
be factorized into quantities measured at $\theta=0$:
\begin{align}
  \avg{H\mathcal{O}}_{\theta}
  &=
  \left(\frac{\calZ_{\theta}}{\calZ_0}\right)^{-1}
  \avg{H}_{0}
  \left\langle
    \mathcal{O}\,
    \rme^{\rmi\theta Z\Qtop^{(\tflow)}[U,B]}
  \right\rangle_{B,0}.
  \label{eq:factorization}
\end{align}
Here $\langle\cdots\rangle_{B,0}$ is the normalized ensemble average with
weight $\exp[-\SW[U,B]]$, and $Z$ is a multiplicative rescaling of the
flowed gluonic topological charge.  The global normalization is
\begin{equation}
  \frac{\calZ_{\theta}}{\calZ_0}
  =
  \left\langle
    \rme^{\rmi\theta Z\Qtop^{(\tflow)}[U,0]}
  \right\rangle_{0}.
  \label{eq:normalization}
\end{equation}
The factor $\avg{H}_0$ is known to obey a perimeter law in the confined
phase and can be evaluated by multistep reweighting
~\cite{deForcrand:2000fi,deForcrand:2005pb}.  It therefore suffices to
determine the long-distance behavior of the last factor in
Eq.~\eqref{eq:factorization}.  At $\theta=2\pi$ we measure
\begin{align}
  X_H(R)
  &=
  \re\left\langle
    \rme^{2\pi\rmi Z\Qtop^{(\tflow)}[U,B]}
  \right\rangle_{B,0},
  \label{eq:XH}
  \\
  X_{HW}(R)
  &=
  \re\left\langle
    W^{(\tflow')}(C)\,
    \rme^{2\pi\rmi Z\Qtop^{(\tflow)}[U,B]}
  \right\rangle_{B,0}.
  \label{eq:XHW}
\end{align}

The numerical obstacle is the phase in Eq.~\eqref{eq:normalization}.
We evolve the links with a $1$-form-covariant DBW2 flow, taking
$c_1=-1.4088$ in the rectangle-improved flow action
~\cite{Luscher:2010iy,QCD-TARO:1999mox,Tanizaki:2024zsu}.
Because $1+12c_1<0$, lattice-scale instantons are stabilized against
collapse, and the topological sectors become sharply resolved at modest
flow time.  We use a tree-level-improved gluonic charge and choose
$Z=Z(\tflow)$ by minimizing the squared distance of
$Z\Qtop^{(\tflow)}[U,0]$ from the nearest integers.  This makes
$\calZ_{2\pi}/\calZ_0$ approach unity and removes a potentially severe
global sign problem.  Crucially, no rounding is applied in the defect
ensemble: when $\rmd B\ne0$, the non-integer values of $\Qtop[U,B]$ are
physical and provide the phase that distinguishes $H$ from $HW$.

\section{Simulation setup}
\label{sec:setup}

We simulate the $SU(2)$ Wilson plaquette action at $\beta=2.45$ on
$12^3\times8$ and $16^3\times10$ lattices.  Both systems are confined at
$\theta=0$.  Heat-bath updates are implemented with JuliaQCD
~\cite{Nagai:2024yaf,bezanson2012juliafastdynamiclanguage}; as a benchmark,
we reproduce the known perimeter law of the 't~Hooft loop at $\theta=0$.
Table~\ref{tab:parameters} lists the ensembles.  The ``$R=0$'' ensembles, which mean the sets of configurations without defect insertion, 
determine $Z$ and the global normalization, while a separate defect
ensemble is generated for every nonzero $R$.

\begin{table}[t]
  \centering
  \small
  \begin{tabular}{@{}cccccc@{}}
    \toprule
    $\beta$ & $\Ns$ & $\Nt$ & $R$ & $\Nconf$ & $N_{\mathrm{skip}}$ \\
    \midrule
    2.45 & 12 & 8  & 0              & 10,000 & 10 \\
         &    &    & $1$, \dots, $12$ & 40,000 & 10 \\
    \midrule
    2.45 & 16 & 10 & 0              & 5,000  & 10 \\
         &    &    & $6$, \dots, $10$ & 80,000 & 10 \\
    \bottomrule
  \end{tabular}
  \caption{Simulation parameters.  Configurations are stored after every
  $N_{\mathrm{skip}}$ heat-bath updates.}
  \label{tab:parameters}
\end{table}

We set $\tflow=0.8$ on the smaller lattice and $\tflow=1.0$ on the larger
one, in lattice units, and use $\tflow'=\tflow$ for the Wilson-loop
component of $X_{HW}$.  Statistical errors are estimated by bootstrap
resampling and include the uncertainty in the rescaling factor $Z$.  The
flow smears an operator over the diffusion length
\begin{equation}
  d=\sqrt{8\tflow}.
  \label{eq:diffusion_length}
\end{equation}
To suppress overlap between the two smeared probes and their periodic
images, we use the window
\begin{equation}
  2d\lesssim R\lesssim \Ns-2d .
  \label{eq:nonoverlap_window}
\end{equation}
For $12^3\times8$, this leaves only a narrow region around the midpoint.
For $16^3\times10$, it gives the useful window $6\le R\le10$, which is why
the second volume is essential.

\section{Numerical results}
\label{sec:results}

\subsection{Topological sectors and global normalization}

\begin{figure}[t]
  \centering
  \includegraphics[width=0.49\linewidth]{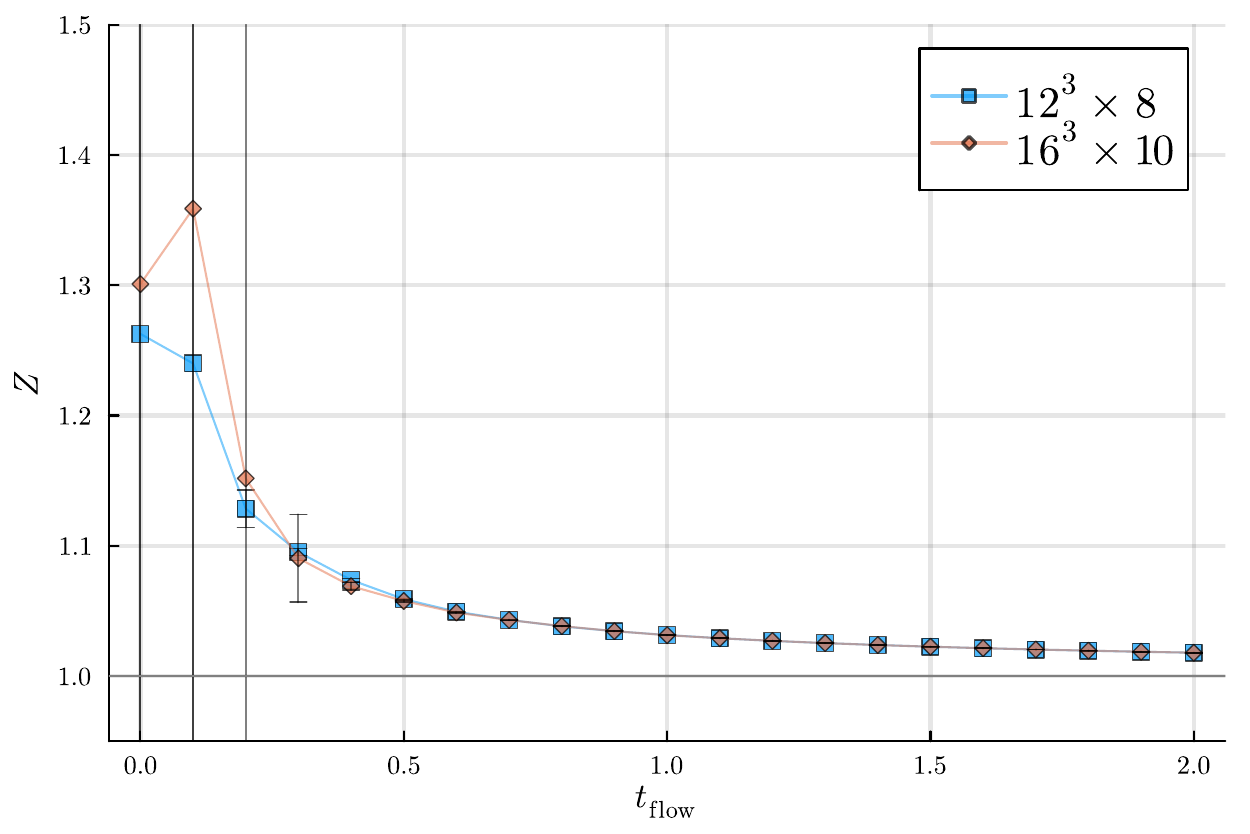}\\
  \includegraphics[width=0.49\linewidth]{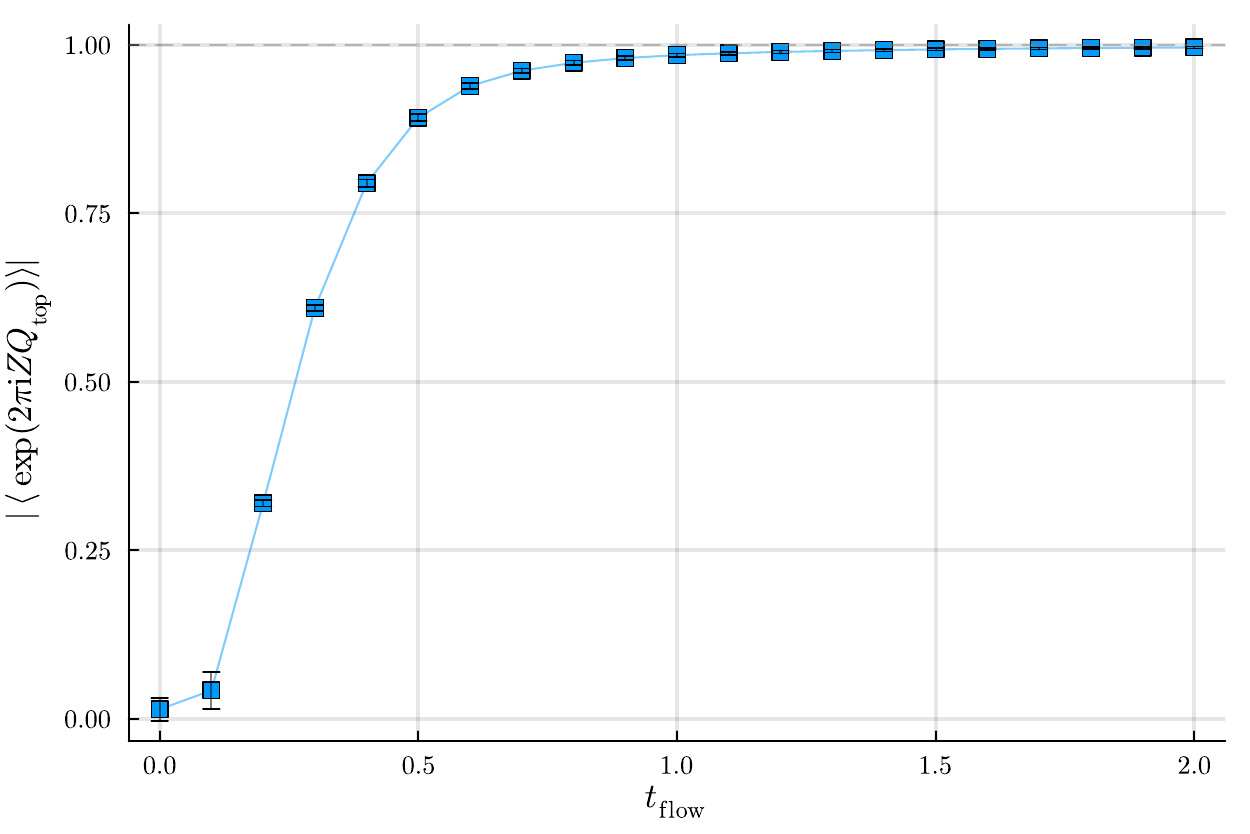}
  \includegraphics[width=0.49\linewidth]{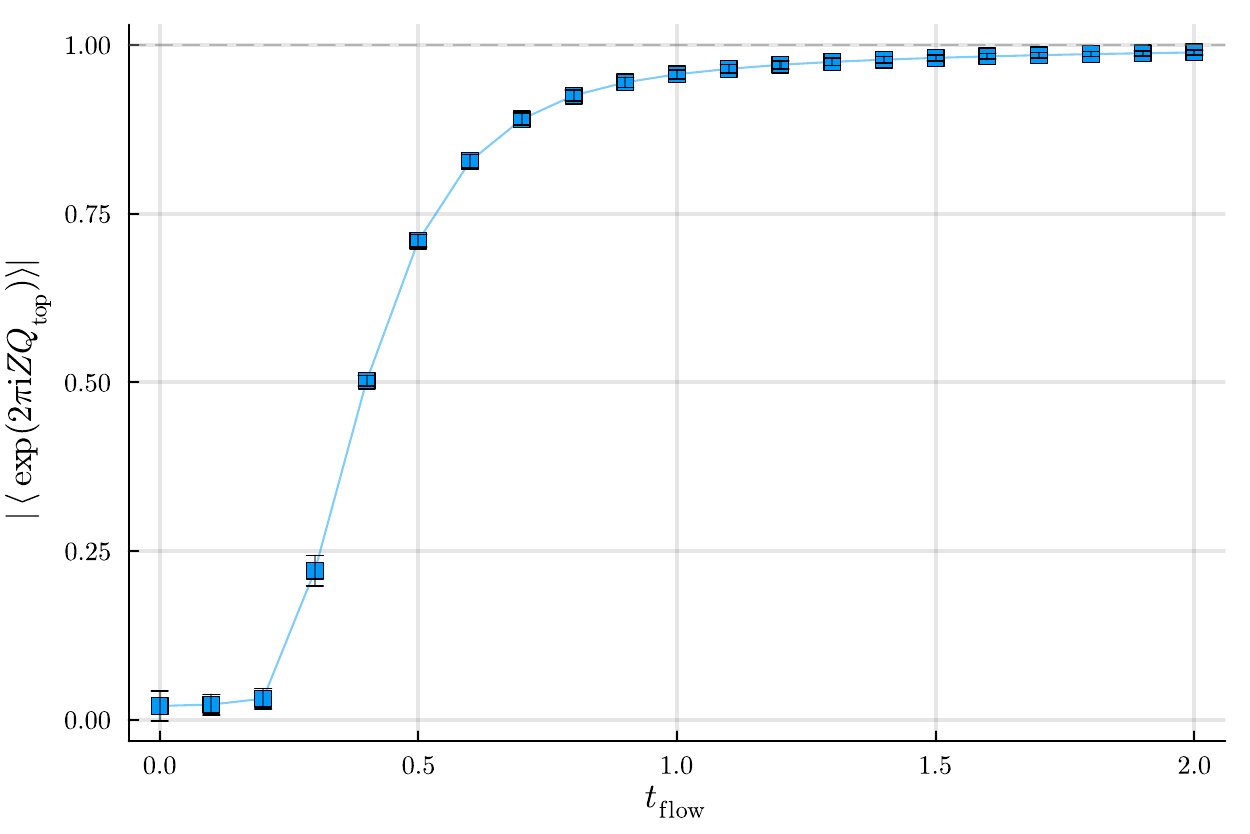}
  \caption{DBW2-flow dependence of the rescaling factor $Z$ (above) and
  of $\lvert\calZ_{2\pi}/\calZ_0\rvert$ on the $12^3\times8$ (below left) and
  $16^3\times10$ (below right) lattices. Reproduced from Ref.~\cite{Watanabe:2026ymb}.}
  \label{fig:normalization}
\end{figure}

Figure~\ref{fig:normalization} shows that $Z$ rapidly approaches a plateau
under DBW2 flow and that the absolute value of the $\theta=2\pi$
normalization approaches unity on both volumes.  The corresponding
histograms develop narrow peaks close to integer values.  The global phase
in Eq.~\eqref{eq:normalization} is therefore numerically benign at the flow
times used below.  By contrast, the defect background remains non-flat and
its topological charge retains continuous fluctuations.

\begin{figure}[t]
  \centering
  \includegraphics[width=0.495\linewidth]{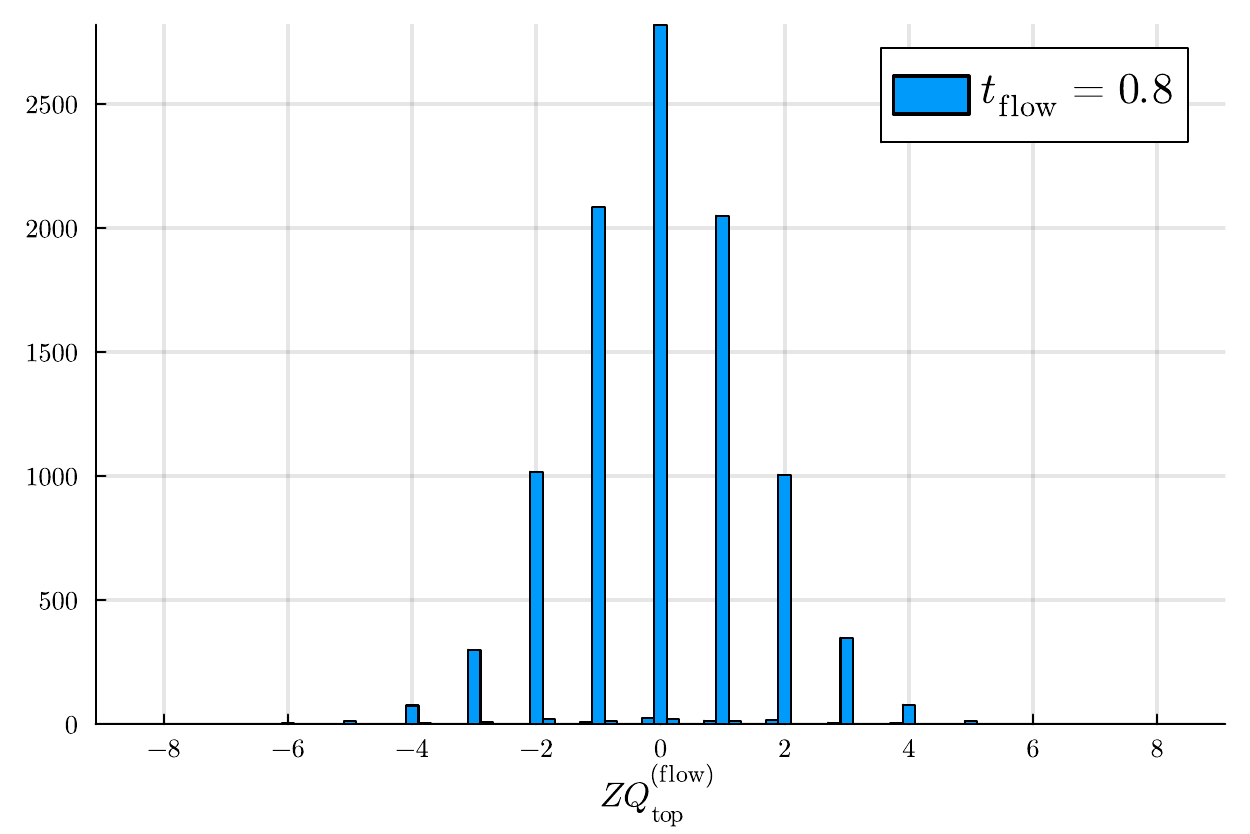}
  \includegraphics[width=0.495\linewidth]{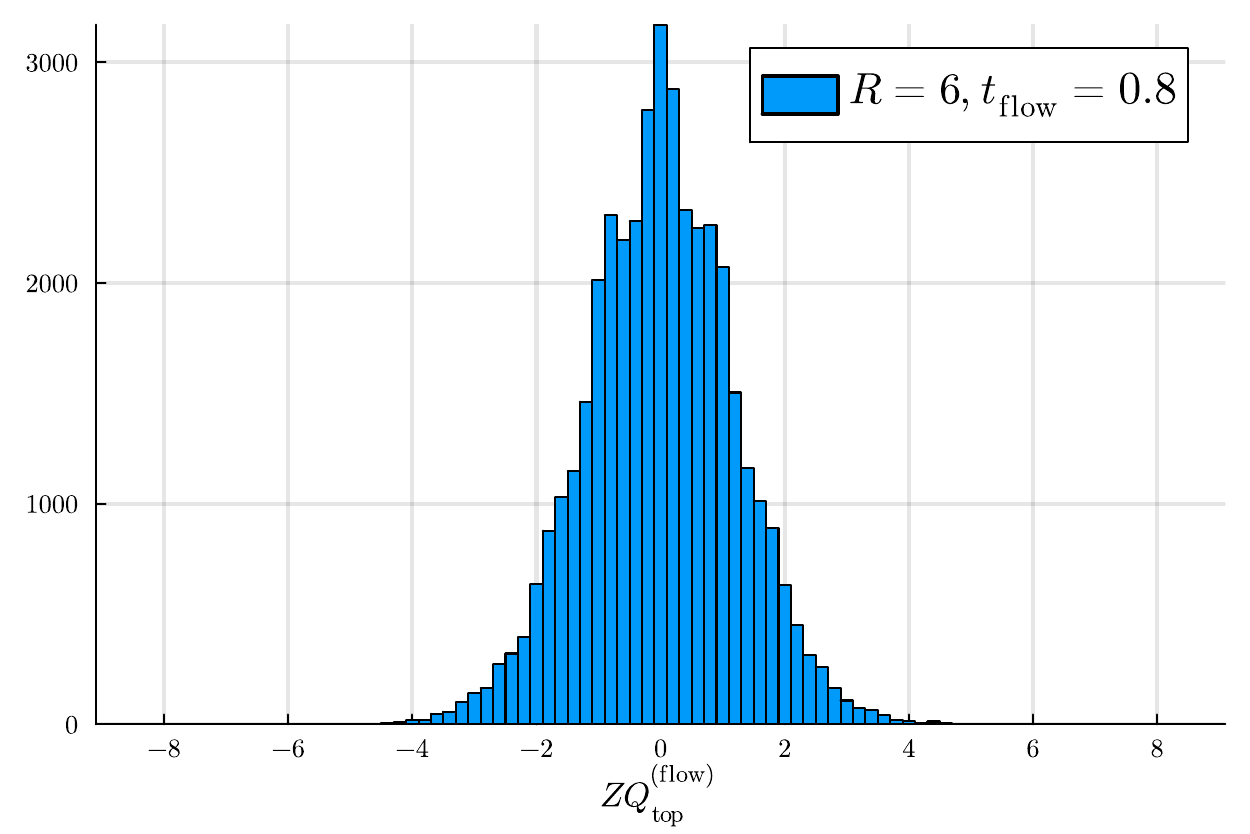}
  \caption{Histograms of the flowed topological charge on the
  $12^3\times8$ lattice at $\tflow=0.8$.  Without a defect ($B=0$; left),
  $Z\Qtop$ resolves into nearly integer sectors.  With an $R=6$
  't~Hooft defect ($\rmd B\ne0$; right), it retains a continuous
  distribution. Reproduced from Ref.~\cite{Watanabe:2026ymb}.}
  \label{fig:histograms}
\end{figure}

Figure~\ref{fig:histograms} makes the contrast explicit on the
$12^3\times8$ lattice.  At $B=0$, the flowed charge is concentrated close
to integer sectors.  In the $R=6$ defect ensemble, where $\rmd B\ne0$, the
distribution remains continuous and must not be rounded.

\clearpage
\subsection{Magnetic line}

\begin{figure}[t]
  \centering
  \includegraphics[width=0.495\linewidth]{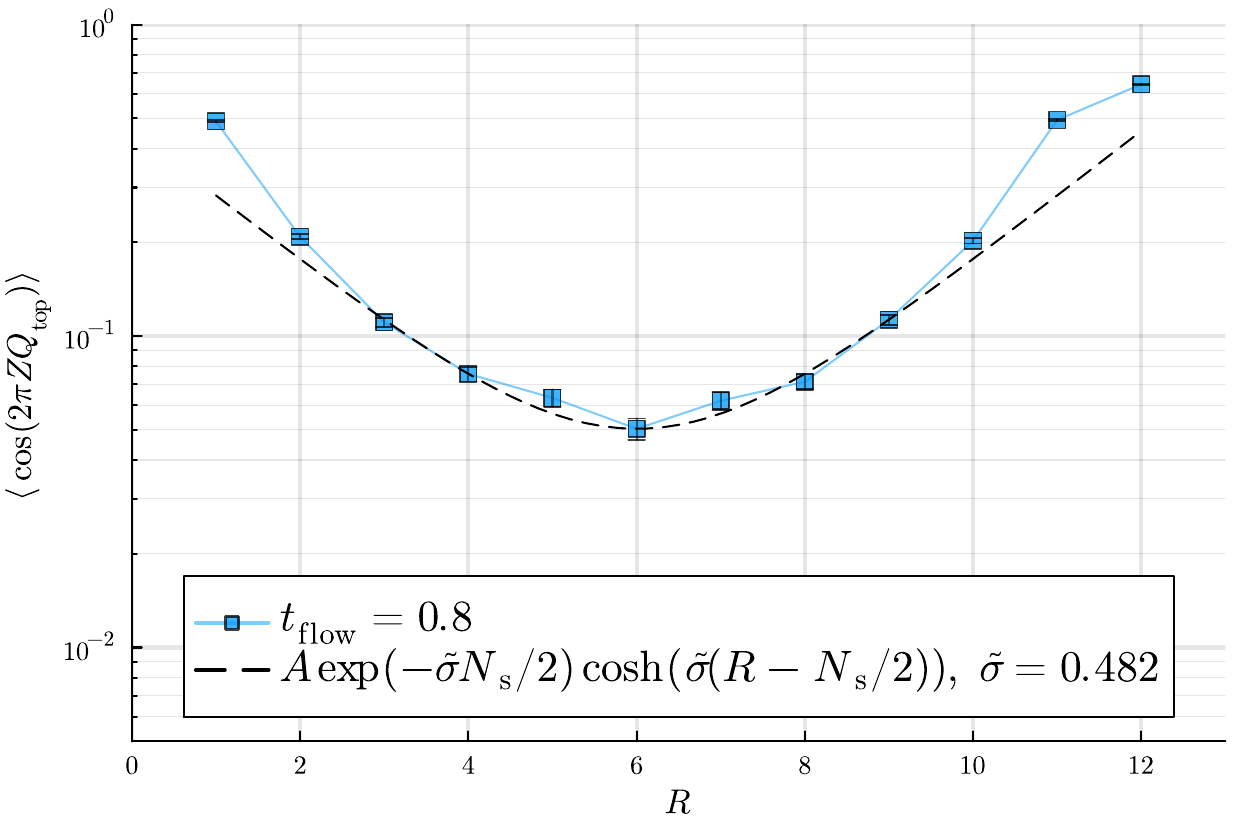}
  \includegraphics[width=0.495\linewidth]{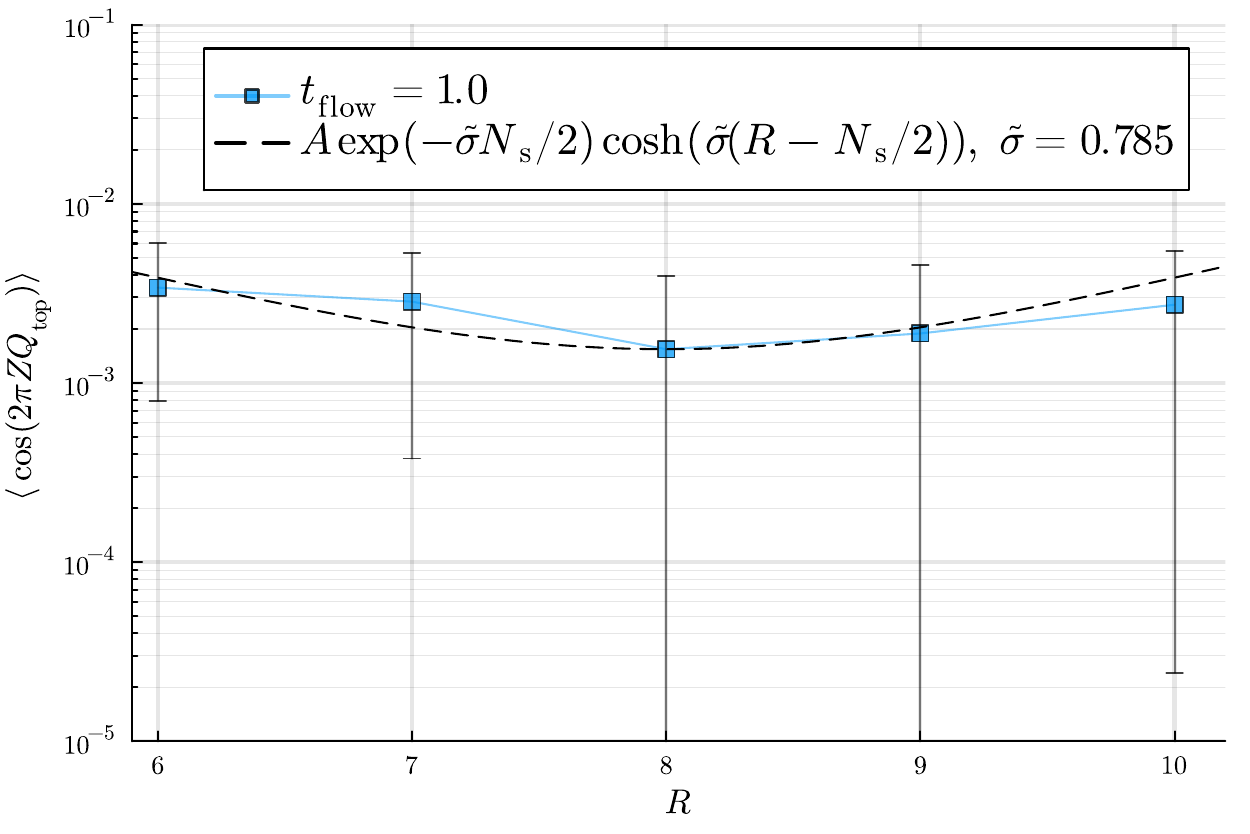}
  \caption{The magnetic determining factor
  $X_H(R)=\avg{\cos(2\pi Z\Qtop)}_{B,0}$ on the
  $12^3\times8$ (left) and $16^3\times10$ (right) lattices.  Dashed
  curves are area-law eye guides based on the independently determined
  Polyakov-loop mass gap. Reproduced from Ref.~\cite{Watanabe:2026ymb}.}
  \label{fig:H_result}
\end{figure}

The real parts of the magnetic determining factor $X_H(R)$ are displayed
in Fig.~\ref{fig:H_result}.  The imaginary parts, which should vanish for
our reflection-symmetric setup, fluctuate around zero within about two
standard deviations and provide a consistency check.  If the magnetic line
obeys an area law,
periodicity in the spatial direction gives
\begin{equation}
  X_H(R)
  \propto
  \rme^{-\widetilde\sigma R}
  +
  \rme^{-\widetilde\sigma(\Ns-R)} .
  \label{eq:area_guide}
\end{equation}
The dashed curves in Fig.~\ref{fig:H_result} are eye guides of this form,
not fits.  The scale $\widetilde\sigma$ is fixed by the fundamental
Polyakov-loop mass gap, and only the overall normalization is matched at
$R=\Ns/2$.

On $12^3\times8$, the signal has a nontrivial $R$ dependence compatible
with Eq.~\eqref{eq:area_guide}, but the non-overlap window in
Eq.~\eqref{eq:nonoverlap_window} is too narrow for a firm conclusion.  On
$16^3\times10$, the central values in the non-overlap window are of order
$1$--$3\times10^{-3}$, comparable to the statistical floor of order
\begin{equation}
  \delta X_H=O(\Nconf^{-1/2}),
  \qquad
  \Nconf^{-1/2}
  =
  (80{,}000)^{-1/2}
  \simeq 3.5\times10^{-3}.
\end{equation}
Their smallness and distance dependence are consistent with an area law,
but the direct signal is noise-limited.  We therefore do not claim an
independent precision determination of the magnetic string tension.

\subsection{Dyonic line}

\begin{figure}[t]
  \centering
  \includegraphics[width=0.495\linewidth]{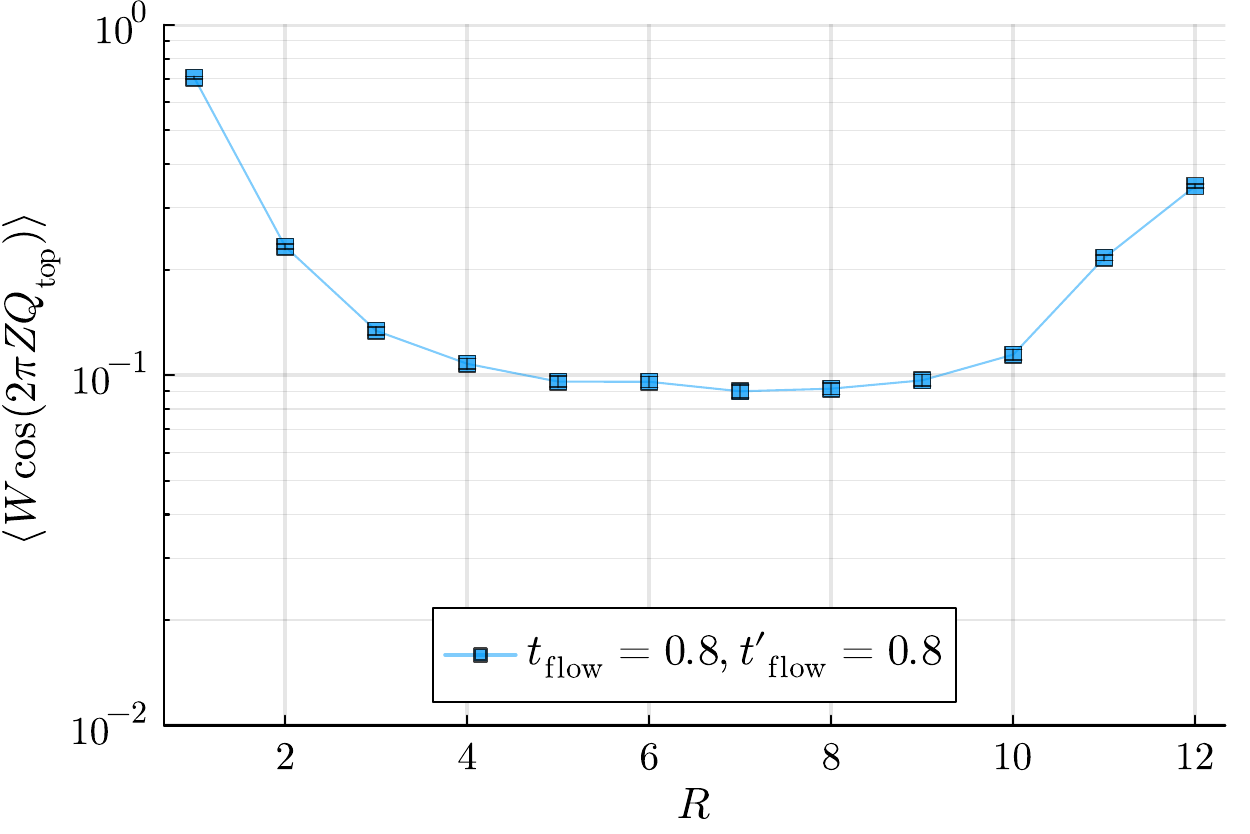}
  \includegraphics[width=0.495\linewidth]{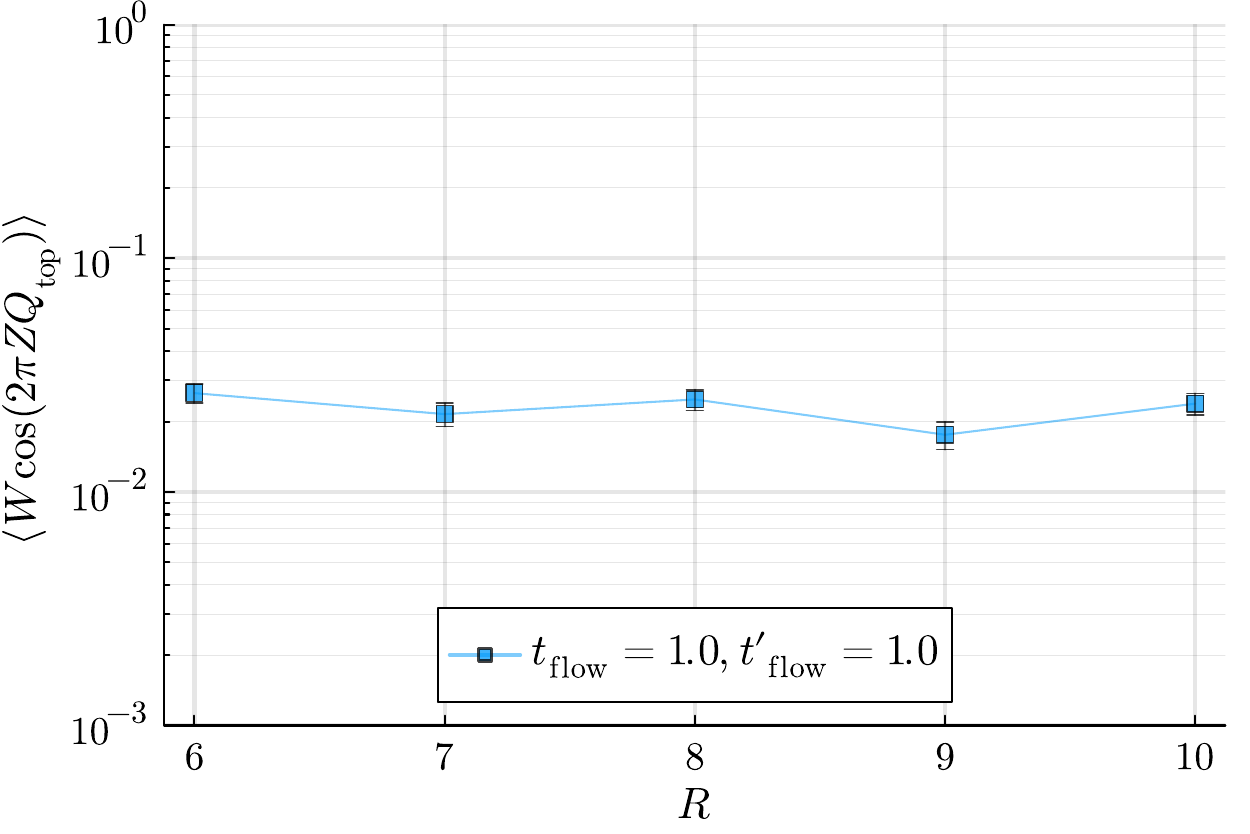}
  \caption{The dyonic determining factor
  $X_{HW}(R)=\avg{W\cos(2\pi Z\Qtop)}_{B,0}$ on the
  $12^3\times8$ (left) and $16^3\times10$ (right) lattices.  The
  long-distance plateaus signal a perimeter law. Reproduced from Ref.~\cite{Watanabe:2026ymb}.}
  \label{fig:HW_result}
\end{figure}

The situation is qualitatively different after dressing the magnetic
defect with a fundamental Wilson loop.  Figure~\ref{fig:HW_result} shows a
weak $R$ dependence for $4\lesssim R\lesssim10$ on $12^3\times8$ and a
plateau throughout the non-overlap window $6\le R\le10$ on
$16^3\times10$.
The signal remains well above the statistical floor.  Thus $X_{HW}$, and
hence the dyonic line, obeys a perimeter law at $\theta=2\pi$.

This difference also explains the contrast in statistical quality.  An
area-law observable becomes exponentially small toward $R=\Ns/2$ and
eventually meets the $O(\Nconf^{-1/2})$ noise floor.  A screened dyonic
line does not suffer this exponential suppression in the same region.
The plateau is stable under moderate changes of $\tflow$, and replacing
the rectangular Wilson contour by a fundamental Polyakov-loop pair on the
$12^3\times8$ ensemble gives the same long-distance behavior.

In a gapped Lorentz-invariant $SU(2)$ vacuum, only one of $W$, $H$, and
$HW$ can have a perimeter law.  The Wilson loop has an area law by
confinement, while Fig.~\ref{fig:HW_result} selects $HW$ as the screened
line.  The Wilson--'t~Hooft classification then implies an area law
for $H$, even though its direct measurement is noise-limited.  The
screened object carries charges $(e,m)=(1,1)$: the condensate is dyonic,
and its electric $\mathbb{Z}_2$ $N$-ality encodes the nontrivial SPT
label expected at $\theta=2\pi$.

\section{Summary and outlook}
\label{sec:summary}

We have tested the SPT distinction between the confining vacua at
$\theta=0$ and $2\pi$ through line operators.  A $1$-form-covariant DBW2
flow makes the global $\theta=2\pi$ reweighting factor close to unity
without rounding the physical topological charge in the defect ensemble.
The dyonic Wilson--'t~Hooft loop shows a clear perimeter-law plateau on
both volumes.  The bare magnetic signal lies at the expected area-law
scale and has compatible distance dependence, although its direct
determination is limited by statistics.  Taken together with the
Wilson--'t~Hooft classification, the data provide numerical evidence that
the screened line changes from $H$ at $\theta=0$ to $HW$ at
$\theta=2\pi$, as predicted by the Witten effect and the $1$-form SPT
shift.

A sharper direct magnetic signal should be possible with an anisotropic
$16\times12^2\times8$ lattice: it retains a wide separation window while
reducing both the defect area and the cost of the gradient flow.  Dyonic
lines built from Polyakov-loop correlators avoid the $R$-dependent
backtracking segment of the rectangular Wilson contour.  It is also
natural to extend the analysis to spatial 't~Hooft and dyonic loops in the
deconfined phase, where their tensions diagnose the corresponding domain
walls~\cite{Hayashi:2026ijm}.

\acknowledgments
This work was triggered by conversations during the workshop
\textquotedblleft Nonperturbative Approaches to QFTs\textquotedblright\
at the Komaba campus of the University of Tokyo in March 2024.  We thank
Yui Hayashi, Zohar Komargodski, Akira Matsumoto, Jun Nishimura, and
Ryutaro Tsuji for useful discussions.  This work was partially supported
by JSPS KAKENHI Grant Numbers JP25K17402 (O.M.), JP20K14479, JP22H05111,
JP22K03539, JP22H05112 (Y.N.\ and A.T.), JP23K22489, JP26K07106 (Y.T.), and
JP24K00630 (H.W.); the Multidisciplinary Cooperative Research Program in
CCS, University of Tsukuba; MEXT's Program for Promoting Researches on
the Supercomputer Fugaku (JPMXP1020230411 and JPMXP1020230409); JST BOOST
JPMJBY24F1 (A.T.); and the Center for Gravitational Physics and Quantum
Information at Yukawa Institute for Theoretical Physics.  The numerical
work used Yukawa-21 at Yukawa Institute for Theoretical Physics and Ajiro
at the Center for Computational Sciences, University of Tsukuba.  O.M.\
acknowledges the RIKEN Special Postdoctoral Researcher Program and RIKEN
FY2025 Incentive Research Projects.

\bibliographystyle{utphys}
\setlength{\bibsep}{2pt}
\bibliography{QFT,ref,coderef,proceedings}

\end{document}